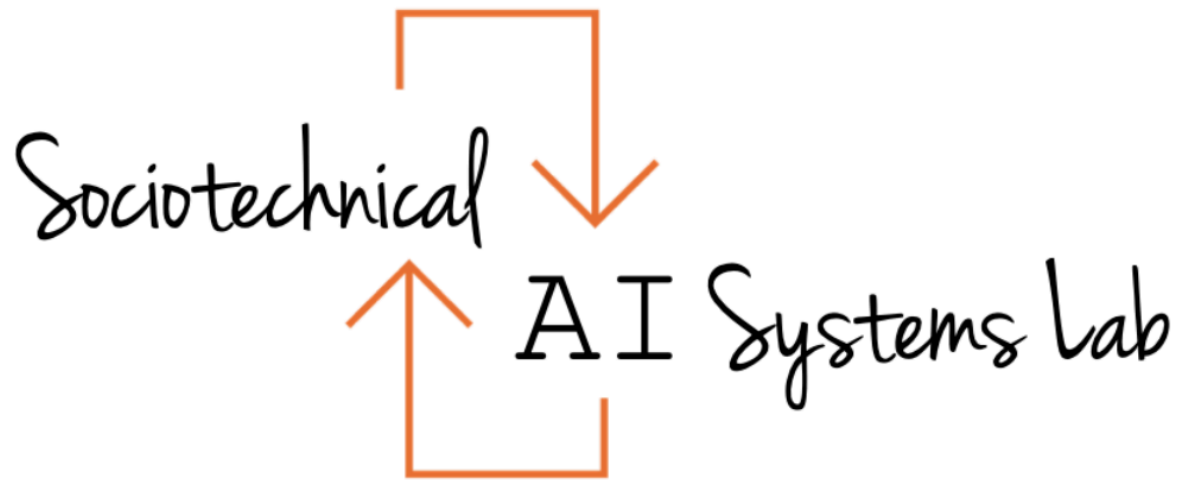

Policy Brief:

AI Safety is Stuck in Technical Terms - A System Safety Response to the International AI Safety Report

Delft, 5 February 2025

By Dr. Roel Dobbe, Delft University of Technology
r.i.j.dobbe@tudelft.nl

# Abstract

Safety has become the central value around which dominant AI governance efforts are being shaped. Recently, this culminated in the publication of the International AI Safety Report, written by 96 experts of which 30 nominated by the Organisation for Economic Co-operation and Development (OECD), the European Union (EU), and the United Nations (UN). The report focuses on the safety risks of general-purpose AI and available technical mitigation approaches. In this response, informed by a *system safety* perspective, I reflect on the key conclusions of the report, identifying fundamental issues in the currently dominant technical framing of AI safety and how this frustrates meaningful discourse and policy efforts to address safety comprehensively. The system safety discipline has dealt with the safety risks of software-based systems for many decades, and understands safety risks in AI systems as *sociotechnical* and requiring consideration of technical and non-technical factors and their interactions. The International AI Safety report does identify the need for system safety approaches. Lessons, concepts and methods from system safety indeed provide an important blueprint for overcoming current shortcomings in technical approaches by integrating rather than adding on non-technical factors and interventions. I conclude with why building a system safety discipline can help us overcome limitations in the European AI Act, as well as how the discipline can help shape sustainable investments into Public Interest AI.

Table of Contents

# 1  Introduction

As we are approaching the AI Action Summit in Paris, I reflect on recent developments in European and global coordination to govern artificial intelligence systems in high-stakes domains. In particular, I respond to *the International AI Safety report* published last week in preparation of the summit.[1] Furthermore, I touch on the current negotiations and formulation of a "General-Purpose AI Code of Practice" to guide the implementation of the European Union's AI Act to address systemic risk.[2] Based on my reflections, I also motivate system safety as a guiding principle for promoting and realizing "Public Interest AI", which is a core aim of the AI Action Summit.[3]

In this report, I will first interpret the key challenges identified in the International AI Safety report, laying out fundamental issues in our current understanding and framing of AI safety and how these frustrate meaningful discourse and policy efforts. I reflect on these issues through the lens of *system safety*, thereby providing a lodestar for informing more integrated AI safety efforts. I conclude with why building a system safety discipline can help us advance AI that is more squarely in the public interest. I end with priorities for the current landscape of actors in industry, civil society and policy/politics, with an emphasis on the European continent.

My response is purposefully shaped as a response to inform the rapidly developing discourse. It surfaces the biggest concerns in the current attempts to institutionalize notions of "safety" in AI governance approaches, as well as the open problems and challenges that lie ahead and how to address them. This response is released in anticipation of the AI Action Summit in Paris during February 10-11, 2025. In the coming weeks, the response may also form input to the last stages of drafting the EU Code of Practice. In the coming months, I may consider expanding this document or providing updates as the regulatory and technological spaces develop.

As such, this response is by no means comprehensive and primarily meant to inspire questions and discussions to improve our thinking and interventions about "AI Safety" and "Public Interest AI". It is not meant to provide a complete account of the value of system safety, but will reference relevant further readings. Furthermore, given the short timeline in which this document was put together, some obvious references are still missing, and will be added in a later edition. Those interested in discussing or sharing issues or references I could add in subsequent writing are invited to reach out to the author.

---

[1] Bengio, Y., Mindermann, S., Privitera, D., Besiroglu, T., Bommasani, R., Casper, S., Choi, Y., Fox, P., Garfinkel, B., Goldfarb, D., Heidari, H., Ho, A., Kapoor, S., Khalatbari, L., Longpre, S., Manning, S., Mavroudis, V., Mazeika, M., Michael, J., … Zeng, Y. (2025). *International AI Safety Report* (arXiv:2501.17805). arXiv. https://doi.org/10.48550/arXiv.2501.17805

[2] *General-Purpose AI Code of Practice | Shaping Europe's digital future*. (n.d.). Retrieved February 5, 2025, from https://digital-strategy.ec.europa.eu/en/policies/ai-code-practice

[3] *Public interest AI*. (2024, October 2). Elysee.Fr. https://www.elysee.fr/en/sommet-pour-l-action-sur-l-ia/public-interest-ai

Before I discuss the International AI Safety Report, I first provide some background on the political backdrop and the need to see AI in light of planetary crises.

## 2  Political Backdrop

For artificial intelligence, the start of 2025 has been nothing short of tumultuous. The inauguration of Trump has triggered hyper investments framing AI as the central source of prosperity in Trump's proclaimed "golden age". President Trump has released a myriad of executive orders attacking regulatory structures in the US that may support the institutionalization of AI safety norms.[4] The deregulatory wind in DC is welcomed by various big tech companies, and inspires those currently in power of AI developments and high-risk AI systems to tread ahead with a markedly lower appetite for risk and regulation. Meanwhile, Chinese investors have released new AI systems that rival the most advanced US systems, both in "performance" on certain benchmarks as well as in terms of the resources needed to train the AI models, thereby questioning the central business model and policy assumptions behind the US's steep AI investments and foreign policy strategies.[5] In Europe, regulators have passed the AI Act last summer, and the start of the year the gaze is set at the development of a code of practice for general-purpose AI systems. This code will be a central anchor and reference point for operationalizing the AI Act by developers and users, and for enforcing it by regulators, and will be released later this Spring.

## 3  AI and Planetary Crises

Meanwhile, a broader understanding is emerging that AI as an industry and technology is not aligned with several alarming planetary crises. Firstly, ecological crises face lower concern and headwinds due to AI policies and renewed investments in fossil fuel developments. Trump's StarGate plan explicitly aims to power additional data centers for AI development with fossil fuels, emblematized by the inaugural mantra, "we will drill, baby, drill". Indeed, it has been known for a longer time that the resource hungry model behind big tech's AI development is presenting extra challenges to addressing planetary crisis, not only those related to climate change but also other concerns.[6] Secondly, inequalities and inequities are bound to deepen, with AI serving as a central device to capture not just the data of people, organizations and institutions, but also the ability to control and shape the future of how we live, educate, regulate and co-exist.[7] Lastly, our political systems in democracy are under attack by those operating AI and other critical technological infrastructures.[8] The aggressive meddling by tech billionaires in elections are a broader sign that wealth and capital are out to consolidate and extend their power by weakening people, societies and democracies. While the developments in DC are felt viscerally by many, dangers to our democratic institutions and rule of law lark broadly across Western societies, and the role of AI and digital technologies is substantial and increasingly understood as a major vector of risk.[9]

---

[4] *Biden's Sweeping AI Order Scrapped by Trump in Regulatory Reset*. (2025, January 21). https://news.bloomberglaw.com/artificial-intelligence/trump-scraps-bidens-sweeping-ai-order-in-regulatory-reset
[5] Friesen, G. (2025). *What DeepSeek's AI Innovation Means For Investors And Big Tech*. Forbes. Retrieved February 5, 2025, from https://www.forbes.com/sites/garthfriesen/2025/01/29/deepseeks-ai-innovation-what-it-means-for-investors-and-big-tech/

As such, one can already conclude that any notion of *safety* as it pertains to artificial intelligence, whether it is understood directly in ensuring safe AI applications, or indirectly in ensuring we safeguard our democratic institutions and planetary life-forms from system risks, seems to be in great peril.

# 4  What is System Safety?

The International AI Safety Report defines system safety as an approach that "applies both engineering and management principles to identify and control hazards throughout a system's life cycle. For general-purpose AI, this includes understanding the interactions between the hardware and software components, organisational structures, and human factors."[10] System safety pioneer Nancy Leveson defines the discipline as "the management of hazards: their identification, evaluation, elimination, and control through analysis, design and management procedures."[11] As such, the discipline integrates knowledge and insight from different actors and disciplines: "[It] includes things such as political and social processes, the interests and attitudes of management, designers or operators, human factors and cognitive psychology, the effects of the legal system on accident investigations and free exchange of information, certification and licensing of critical employees and systems, and public sentiment."[12]

---

[6] Dobbe, R., & Whittaker, M. (2019). *AI and Climate Change: How they're connected, and what we can do about it*. AI Now Institute, New York University. https://ainowinstitute.org/publication/ai-and-climate-change-how-theyre-connected-and-what-we-can-do-about-it

[7] Whittaker, M. (2021). The steep cost of capture. *Interactions*, *28*(6), 50–55. https://doi.org/10.1145/3488666

[8] Schaake, M. (2024). *The Tech Coup: How to Save Democracy from Silicon Valley*. Princeton University Press. https://doi.org/10.1515/9780691241180

[9] Greenstein, S. (2022). Preserving the rule of law in the era of artificial intelligence (AI). *Artificial Intelligence and Law*, *30*(3), 291–323. https://doi.org/10.1007/s10506-021-09294-4

[10] Footnote 1, page 158.

[11] Leveson, N. G. (2023). *An Introduction to System Safety Engineering*. MIT Press. https://books.google.com/books?hl=de&lr=&id=JrKmEAAAQBAJ&oi=fnd&pg=PR13&dq=nancy+leveson+introduction&ots=dcCNz3bOEF&sig=2sqRoTRAed9E5NMfRtb55RuQTQs

As such, the system safety tradition bares in it many lessons, as well as an integrative understanding of the various factors and interventions relevant to understand and manage safety in software-based systems. For a comprehensive introduction to system safety, the reader can consider Leveson's textbooks[13] or read how Leveson's lessons are relevant to safety challenges related to artificial intelligence systems[14].

# 5 Key Conclusions from the International AI Safety Report

The broad lens on planetary crises is reflected in the scope of the International AI Safety (IAIS) Report, consisting of nearly 300 pages, published by "a diverse group of 96 Artificial Intelligence (AI) experts contributed to this first full report, including an international Expert Advisory Panel nominated by 30 countries, the Organisation for Economic Co-operation and Development (OECD), the European Union (EU), and the United Nations (UN)."

The IAIS Report's main conclusions are quite comprehensive and informative in pointing out fundamental issues in the dominant approaches to AI Safety today. I will first cover the key concerns across three areas of the report, to then respond to these in the following section.

The first section concerns the "capabilities" ascribed to AI systems, or what AI "can and cannot do". The report states that it is currently still challenging "to reliably estimate and describe the capabilities of general-purpose AI" with *technical* means. Relatedly, there is broad disagreement on both the potential for progress in capabilities as well as in the nature, emergence and intensification of risks. It further observes that leading companies are betting on scaling" as key to the development of "capabilities", with annually a 4x increase in the use of computational resources, which amounts to a 10,000x increase from 2023 to 2030 if current trends continue.

---

[12] Idem.
[13] Idem; Leveson, N. G. (2011). *Engineering a safer world: Systems thinking applied to safety*. The MIT Press. https://library.oapen.org/handle/20.500.12657/26043

[14] Dobbe, R. (2022). System Safety and Artificial Intelligence. In J. B. Bullock, Y.-C. Chen, J. Himmelreich, V. M. Hudson, A. Korinek, M. M. Young, & B. Zhang (Eds.), *The Oxford Handbook of AI Governance* (p. 0). Oxford University Press.
https://research.tudelft.nl/en/publications/system-safety-and-artificial-intelligence

The second section addresses the risks considered by the authors, which are grouped into malicious use risks, risks from malfunctions, and systemic risks. Malicious use risks emerge when general-purpose AI is used to cause harm to individuals, organizations, or society, and include harm to individuals through fake content, manipulation of public opinion, cyber offense, and biological and chemical attacks. Risks from malfunctions are understood as able to lead to "unintended harm" include reliability issues, bias, and loss of control. Systemic risks are risks associated with "widespread deployment of general-purpose AI" and include labour market risks, global AI R&D divide, market concentration and single points of failure, environmental risks, privacy risks, and copyright infringements.

The various risks are treated without broader reflection on how complete these lists are, and what process was taken to determine and negotiate what risks would be included in the current report. In the next section I will reflect on these aspects - What could be missing and why? And what is needed to arrive at an inclusive list? - rather than discussing individual risks.

The third and last section of the report covers risk mitigation approaches. Firstly, it is crucial to point out that the report focuses primarily on *technical* approaches to risk mitigation. It concludes that "[s]everal technical approaches can help manage risks, but in many cases the best available approaches still have highly significant limitations and no quantitative risk estimation or guarantees that are available in other safety-critical domains." As we'll see in the next section, this conclusion is not surprising and should alarm policy makers and practitioners that a different framing of safety and risk analysis and management is now broadly warranted. The report rightfully points out some distinctive challenges to risk mitigation emanating from features of general-purpose AI. These include the unusually broad range of possible uses and use contexts for general-purpose AI systems, the lack of understanding among developers of the way AI models operate, and the higher level of autonomy of AI systems in acting, planning and delegating tasks.

Moving beyond technical factors, the report highlights the importance of non-technical factors across, focusing on the lack of evidence in policy making in addressing rapidly emerging risks due to depending on general-purpose AI systems, the information gap between what AI companies know about their AI systems and what governments and non-industry researchers know, and the importance of competitive pressures in deprioritizing risk management.

Notably, the report concludes that existing technical risk assessments are severely limited, often "often miss[ing] hazards and overestimate or underestimate general-purpose AI capabilities and risks, because test conditions differ from the real world." Such insights are no surprise as technical measures are typically not intended to address contextual factors. Concerningly, current technical methods cannot reliably prevent even "overtly unsafe outputs" of AI models. Reliability is thus a gaping open issue. However, as we'll discuss in the next section, (AI) reliability is related to but not the same as (AI) safety.

A last observation from the third section regards the need for 'system safety' approaches, which is relevant to the central perspective of this response. At the top of the third section, system safety approaches are pointed to for integrally addressing risks across engineering and organizational interventions. This insight does not feature in the executive summary and recommendations of the report and deserves broader attention. However, the acknowledgment of system safety approaches is encouraging as this hold promise to offer a plethora of lessons, concepts and methods to help actualize more effective risk analyses and elimination or mitigation approaches across different practices and practitioners.

# 6 Fundamental Issues in the Technical AI Safety Discourse

In this section, we cover common issues and misconceptions in the current (technical) AI Safety discourse. Doing so, we contrast the dominant understanding of safety with lessons and insights from the system safety discipline, primarily leaning on the work of Professor Nancy Leveson[15] and recent interpretation of these lessons for artificial intelligence[16] and AI governance[17].

---

[15] Leveson, N. G. (2011). *Engineering a safer world: Systems thinking applied to safety*. The MIT Press. https://library.oapen.org/handle/20.500.12657/26043; Leveson, N. G. (2023). *An Introduction to System Safety Engineering*. MIT Press.

[16] Dobbe, R. (2022). System Safety and Artificial Intelligence. In J. B. Bullock, Y.-C. Chen, J. Himmelreich, V. M. Hudson, A. Korinek, M. M. Young, & B. Zhang (Eds.), *The Oxford Handbook of AI Governance* (p. 0). Oxford University Press.
https://research.tudelft.nl/en/publications/system-safety-and-artificial-intelligence

[17] Delfos, J., Zuiderwijk, A. M. G., van Cranenburgh, S., Chorus, C. G., & Dobbe, R. I. J. (2024). Integral system safety for machine learning in the public sector: An empirical account. *Government Information Quarterly*, *41*(3), 101963 https://www.sciencedirect.com/science/article/pii/S0740624X24000558; Nouws, S. J. J., & Dobbe, R. I. J. (2024). The Rule of Law for Artificial Intelligence in Public Administration: A System Safety Perspective. In *Digital Governance: Confronting the challenges posed by Artificial Intelligence*. TMC Asser Press. https://surfdrive.surf.nl/files/index.php/s/gjt6Cg8RgxEVpWE; Rismani, S., Dobbe, R., & Moon, Aj. (2024). *From Silos to Systems: Process-Oriented Hazard Analysis for AI Systems* (arXiv:2410.22526). arXiv. https://doi.org/10.48550/arXiv.2410.22526

## 6.1 The dominant AI safety discourse is stuck in technical terms

Both AI systems and safety, a core value to understand and manage for AI systems, cannot be understood in purely technical terms. This understanding is partly acknowledged by the International AI Safety Report, but otherwise ignored in its core focus and framing of AI safety as a problem requiring technical risk mitigation solutions.

Given that advances of (general-purpose) AI models and applications are shaped by a dominance of scholars and practitioners from software engineering and computer science, it is also understandable that many of the efforts to address safety are informed by the technical concepts and tools that are natural to these disciplines. This line of reasoning can be further supported by taking a longer view of history. Computer systems and software-based automation and decision-making have emerged more dominantly across society over the past 30 years. While it is fair to say that the broad adoption of safety management and risk mitigation practices for applications subject to software-based forms of automation is long overdue, it is not surprising that these are not yet broadly embraced. The establishment and maturation of effective and broadly supported safety knowledge, norms, standards and practices for newly emerging technological applications has often taken many decades, often facing many headwinds even after norms are well-established.

## 6.2 Both AI and safety cannot be properly described in technical terms

The central understanding in safety sciences, in particular those concerned with software-based safety-critical systems, is that safety is a property that can only be understood at the level of a system consisting of different things that act together as a whole to achieve some common goal, objective or end.[18] The 'things' in a system include technical components such as an AI model, the software running the model and an interface to interact with a model, as well as non-technical parts including human beings such as operators, users or subjects as well as broader organizational objects such as protocols or work instructions. Safety is a *system property* which can only be determined by examining the behavior of all the components working together along with the environment in which the components are operating. System behaviors emerging from interactions between technical and social components are therefore best understood as *sociotechnical*.

---

[18] Leveson, N. G. (2023). *An Introduction to System Safety Engineering*. MIT Press.

Using this understanding, we can identify two core issues in the understanding of 'AI system' and 'safety' in the IAIS Report. First, surprisingly, the definition of an "AI System" is not provided in the Glossary. The closest glossary term included is: "System: An integrated setup that combines one or more AI models with other components, such as user interfaces or content filters, to produce an application that users can interact with." This definition acknowledges the modular nature of AI systems, while not including the user and other non-technical components. The definition is best understood as a *technical subsystem* of the broader AI system, which includes non-technical elements. As such, the definition is too narrow to properly understand how unsafe system behaviors emerge, e.g. in the interaction between user and the interface and AI model, or by connecting an AI interface to other technical systems such as online markets or services. While various risks may benefit from interventions at the level of the technical subsystem (model, filters, interface), establishing this definition implies a limited system boundary that excludes other possible interventions that address parts of the AI system or interactions therewith (such as human factors, protocols, bans, training requirements).

Second, and perhaps less surprisingly, the definition of safety is primarily informed by the technical scope focusing on model outputs: "Safety (of an AI system): The property of avoiding harmful outputs, such as providing dangerous information to users, being used for nefarious purposes, or having costly malfunctions in high-stakes settings." However, we also read aspects in this definition that point to the broader system nature of safety, namely in the "use for nefarious purposes" or "costly malfunctions in high-stake settings". As such, this definition does not embrace the sociotechnical system nature of safety, but does provide an opening to rethink AI safety in such terms.

## 6.3 AI Safety is misunderstood as model reliability

We now consider a core issue that keeps emerging throughout the history of system safety. In her magnum opus, Leveson centers this as one of the most sticky misconceptions in safety discussions and practices: equating reliability with safety. Let's consider this conclusion in the IAIS Report: "There has been progress in training general-purpose AI models to function more safely, but no current method can reliably prevent even overtly unsafe actions."[19] While the issue is subtle, it is important. What this statement shows us is (1) an emphasis on models, and (2) a focus on reliability. The methods considered in the report are *technical* risk mitigation or *technical* AI safety methods addressing *reliability* of model outputs. In contrast to safety, reliability *can* be understood as a component property and therefore analyzed at the level of an AI model. Surely, the hope is that reliability is a good proxy for safety, but technical methods do not consider safety, which is a property of the broader system in which an AI model operates and interacts.

---

[19] Footnote 1, page 191.

Following Leveson's core lessons, we can make two core statement:

1. *High reliability is not sufficient for safety*: Leveson argues that improving reliability used to be a good proxy for safety when most accidents involved (technical) component failures. However, that is no longer true and reliability and safety can even be conflicting. Many accidents can occur in systems with perfectly reliable components. An example of this is the rise of increasingly more accurate and reliable facial recognition systems. In the hands of those with malicious intent, such systems may work perfectly to reliably marginalize certain individuals or groups, leading to grave safety concerns and harm.

2. *High reliability is not necessary for safety*: this lesson may in fact be a blessing in disguise for some of the reliability issues that we still face. We have many technological systems with components that are not perfectly reliable, but which function safely nonetheless. Consider for instance the use of AI models for weather prediction in aviation. Wrong predictions may be costly but are typically acceptable, as these can be updated with real-time measurements and pilots are trained to respond to sudden/unanticipated weather changes. Put differently, as long as the broader sociotechnical system is designed properly to prevent unreliable AI outputs from leading to actual hazards, safety may be guaranteed.

## 6.4 Foreseeable malfunctioning should not be framed as cause for 'unintended harms'

The report emphasizes the risk of unintended harm by stating that "[e]ven when users have no intention to cause harm, serious risks can arise due to the malfunctioning of general-purpose AI,"[20] through reliability issues, bias or loss of control. However, the text remains unclear about whose intentions and responsibilities matter when it comes to risks of malfunctioning general-purpose AI.

The group centered in this part of the report are the users that use a malfunctioning general-purpose AI that leads to harms that they did not foresee. It is often unrealistic to expect that laymen users of a systems will spot inherent malfunctioning so to prevent risks of downstream harm. This is a responsibility that should not be delegate to laymen users in safety-critical contexts.

---

[20] Footnote 1, pages 18, 88-91.

Once we know malfunctioning risks are emerging in the context of use through both technical and human factors, they may be unclear to users, but they become foreseeable for system developers. Therefore the responsibilities for preventing risks due to such malfunctioning should be an integral part of the actors developing or deploying systems.

The framing of the report around the intentions of "users" is therefore somewhat narrow and could be read as a delegation of responsibility by actors who may easily foresee and prevent such issues to users who lack a reasonable environment or knowledge to prevent such risks, a phenomena sometimes referred to as *moral crumple zones*.[21] As such, a richer sociotechnical understanding of malfunctioning and its associated hazards is crucial, so to assign proper responsibility and accountability.

## 6.5 Sociotechnical AI systems are more than the sum of their (technical and social) parts

A recurring issue in safety science is that engineers and social scientists focus on either the social aspects or the technical aspects of systems and not their interaction. This divide can be observed in the IAIS Report, as well as for the EU General-Purpose AI Code of Practice, in which technical and non-technical aspects and approaches are split across different working groups, and consequently split in the current draft of the code. Improving safety requires focusing on the sociotechnical system as a whole and designing interventions across the social and technical parts of a system and its interactions. While this is not straightforward, many successful examples exist.

## 6.6 Safety can only be designed into/across systems, never solely into AI models

The current technical AI safety ecosystem relies heavily on post hoc approaches that take the "foundation model" as a given and 'add on' additional software components to tweak and filter the outputs of a broader set of components, for instance through applying feedback techniques such as Reinforcement Learning from Human Feedback.[22] While these techniques on their own might contribute to the *reliability* of an AI model, these cannot be used by themselves to claim *safety*. Since safety is a system property, one needs to design for safety at the broader system level and translate safety requirements that resemble what is needed in the context of use top-down to the requirements of the components and interactions of components in the system.

---

[21] Elish, M. C. (2019). Moral crumple zones: Cautionary tales in human-robot interaction (pre-print). *Engaging Science, Technology, and Society*. https://papers.ssrn.com/sol3/papers.cfm?abstract_id=3352057

[22] Lindström, A. D., Methnani, L., Krause, L., Ericson, P., Troya, Í. M. de R. de, Mollo, D. C., & Dobbe, R. (2024). *AI Alignment through Reinforcement Learning from Human Feedback? Contradictions and Limitations* (arXiv:2406.18346). arXiv. https://doi.org/10.48550/arXiv.2406.18346

## 6.7 General-purpose AI subjects us to the 'curse of flexibility'

The curse of flexibility refers to the dominant assumption that, in contrast to physical machines, software is not subject to physical constraints, leading to ever more flexible and complex designs aiming to address existing and emerging requirements. When issues with software-based systems emerge in practice, it often leads to patching additional layers of software. Typically, competitive or financial pressure may inform prioritizing such patching over revisiting the design of the broader system to understand the source of the issue and redesigning the software and system accordingly.

Traditionally, engineers were subject to the structural integrity of the materials they had to work with. Leveson argues that the limiting factors of software-based systems change from the structural integrity of materials to limits on people's intellectual capabilities. Leveson argues that "[t]he most serious problems arise [..] when nobody understands what the software should do or even what it should not do."[23] Large generative AI models are broadly understood as inherently inscrutable. The IAIS report confirms that, while some progress is made, we are far removed from understanding the internal characteristics of large AI models and their neural architectures.[24] Through the lens of software and system safety, Generative AI can thus be seen as the epitome of the curse of flexibility.

If we adopt this conclusion, the implications for safety are concerning. The hope of companies building and hosting "foundation models" is that many subsequent "AI products" are built on top of it. However, most serious problems with software relate to outsourcing software development, as this creates extra distance between those programming the software and those understanding the physical and social realization of the system (domain experts, users, affected stakeholders), thereby also diminishing the autonomy of those stakeholders to remain control over the functionality of vital systems and process.

---

[23] Leveson, N. G. (1995). Safeware: System safety and computers. *ACM New York*. https://scholar.google.com/scholar?cluster=13047022551994532699&hl=en&oi=scholarr
[24] Footnote 1, page 22.

## 6.8 Bigger is not better, blind belief in scaling contributes to safety hazards and planetary crises

In the current paradigm, the size and further scaling of AI is often equated with "better" models, while various AI scholars have warned longer that this assumption is fundamentally flawed[25] and contributes to large environmental consequences[26] as well as damage to cultural heritage and harm to groups and individuals subject to an LLM's errors[27]. The IAIS report does mention some of these concerns but also centers the role of scaling as potentially pivotal to further progress in AI, while acknowledging the existence of disagreement among experts.

The recently celebrated Deepseek models, while still astronomical in size, are believed to require significantly less compute resources to achieve functionality that is state-of-the-art as compared to current benchmarks. Part of the Deepseek story revolves around the possible lack of more powerful GPU chip technology available in China due to trade restrictions, forcing developers to find creative ways to perform better with less powerful chips. While the details are still debated, a central feature of this success story is the more clever use of constraints, composition and modularity in the design of the (technical) AI model. While all safety concerns remain for these new architectural designs, the prominence of system design will definitely aid the inspiration of novel AI designs that are smaller scale and potentially more amenable to safety. However, given the inherent inscrutability of these models, overcoming the curse of flexibility will remain an elusive goal.

What is easier to predict is that the drive towards efficiency will not necessarily reduce the overall use of compute resources, energy use and climate implications. The rebound effect or Jevon's paradox states that AI systems are bound to be used more at large once efficiency drives down the price, increasing competition and access.[28]

---

[25] Ullrich, K., Meeds, E., & Welling, M. (2017). *Soft Weight-Sharing for Neural Network Compression* (arXiv:1702.04008). arXiv. https://doi.org/10.48550/arXiv.1702.04008
Varoquaux, G., Luccioni, A. S., & Whittaker, M. (2024). *Hype, Sustainability, and the Price of the Bigger-is-Better Paradigm in AI* (arXiv:2409.14160). arXiv. https://doi.org/10.48550/arXiv.2409.14160
[26] Dobbe, R., & Whittaker, M. (2019). *AI and Climate Change: How they're connected, and what we can do about it*. AI Now Institute, New York University. https://ainowinstitute.org/publication/ai-and-climate-change-how-theyre-connected-and-what-we-can-do-about-it
[27] Bender, E. M., Gebru, T., McMillan-Major, A., & Shmitchell, S. (2021). On the Dangers of Stochastic Parrots: Can Language Models Be Too Big? 🦜. *Proceedings of the 2021 ACM Conference on Fairness, Accountability, and Transparency*, 610–623. https://doi.org/10.1145/3442188.3445922
[28] Bhardwaj, E., Alexander, R., & Becker, C. (2025). *Limits to AI Growth: The Ecological and Social Consequences of Scaling* (arXiv:2501.17980). arXiv. https://doi.org/10.48550/arXiv.2501.17980

Furthermore, the deep belief in scale pervasive in Silicon Valley is antithetical to addressing systemic issues related to marginalization and inequality.[29]

## 6.9 Post hoc technical mitigation techniques may be too little too late

As a result of the above, the heavy emphasis on post hoc technical risk mitigation for AI safety can easily be misguided, and safety practitioners should consider two things. First, it is important to realize that "mitigation" inherently means that we accept that certain risks are going to remain, and we do our best to reduce the likelihood or possible impact of these risks leading to concrete hazards and possible accidents and harm. This assumption may be warranted but given the problematic nature of overly flexible and inscrutable models, this also means that ruling out unsafe outputs at the level of the model is an unachievable challenge. As a result, subsequent mitigation techniques should either focus on whether there are acceptable ways to work with the presence of such unsafe outputs, by ensuring these do not contribute to the violation of downstream safety norms. Or by creating the factual basis on which one may conclude that such outputs are fundamentally unsafe, and the use of the AI model needs to be reconsidered.

Second, safety goals and norms can only be expressed at the broader system level and from there translated to requirements on the AI model. In absence of such contextualization, we cannot make any claims on safety. This leads us to a final conclusion of this response.

# 7 Conclusion: "Safe General-Purpose AI" is an oxymoron

The unsatisfying conclusion of the IAIS report is that the best available technical mitigation techniques, typically addressing the model or adjacent technical components of an AI system, still have highly significant limitations. Our interpretation of the findings in the report shows that this conclusion is not surprising and necessitates a pivot in how we think collectively about AI and safety.

A crucial corollary is that AI models cannot be safe in isolation. As Leveson argues, claims about the safety of a piece of software are meaningless without consideration of the broader context in which it operates.

---

[29] Hanna, A., & Park, T. M. (2020). *Against Scale: Provocations and Resistances to Scale Thinking* (arXiv:2010.08850). arXiv. https://doi.org/10.48550/arXiv.2010.08850

It is promising that the need for system safety is now more broadly seen and acknowledged in the AI safety discourse. The lessons from system safety, partly reflected in this response, but rigorously laid out by Professor Nancy Leveson, can help accelerate a more accurate language and understanding of safety that can overcome the limitations of current paradigms and approaches, thereby stimulating more effective forms of knowledge.

Does this mean that General-Purpose AI Systems are fundamentally unsafe? In some ways, yes. The limits of what can be done with AI are different than what can be done safely. Pushing systems to be more general-purpose means pushing systems into higher complexity domains which come with inherent sources of hazard. The history of software and safety provides important insights into this tradeoff.

This should give those wanting to build and depend on general-purpose AI in safety-critical contexts pause. Lessons about the ways in which overly complex and inscrutable software contribute to safety hazards and harms should not be overlooked and be central considerations in the analysis and design of safety-critical AI systems. Accordingly, the methods available to understand AI models in their sociotechnical systems context should be leading in informing what limits and interventions are needed once we do use such AI models.[30]

So how "general-purpose" should future AI systems in safety-critical context be? For many safety-critical context, the more logical conclusion may be to not use general-purpose AI and build lower complexity models or other solutions that are more amenable to systems engineering, thereby allowing for system design interventions that help attain safety norms.

---

[30] Rismani, S., Dobbe, R., & Moon, Aj. (2024). *From Silos to Systems: Process-Oriented Hazard Analysis for AI Systems* (arXiv:2410.22526). arXiv. https://doi.org/10.48550/arXiv.2410.22526

# 8  What System Safety Could Mean for the EU Code of Practice

The term "AI Safety" has taken flight within the policy domain, most starkly since the first AI Safety Summit in Bletchley Park November 2023. At that time, the EU's draft AI Act had already converged around a focus on safety. While the EU Act centers a broad spectrum of public values, including fundamental rights and environmental implications, its legal articles are centered around adherence to harmonized product safety standards. The emergence of safety as the central value in technology development is not new and echoes the emergence of regulation for many previous technological advances. The need for a broader spectrum of values and rights to be safeguarded around the development and use of AI systems is however well documented. That said, safety was agreed upon as the central guiding value or principle through extensive negotiation across EU member states and with the consultation and input from a wide variety of stakeholders from industry and civil society, be it spectacularly lobbied by non-European tech companies.[31]

The more narrow focus on product safety in the AI Act in forming the bedrock against which enforcement is possible has been thoroughly criticized. In particular, there are broadly shared concerns that a risk-based approach undermines the AI Act's ability to provide effective protection for human rights in the context of AI systems.[32]

More recently, the Corporate Europe Conservatory showed how Europe's standard-setting bodies dealing with AI are heavily skewed towards the tech industry. Attempts at broader participation are unsuccessful in preventing corporations from writing rules that will govern their own AI products, "including on how to prevent fundamental rights violations of European citizens."[33]

My reflections on the International AI Safety Report echo these concerns. Product safety may form part of the response to safety, but most AI safety risks as well as concerns around rights and environmental implications require a broader sociotechnical system understanding.[34] Big tech companies would benefit from one- or some-size-fits-all standard approaches to certify their general-purpose AI systems. The lessons from system safety however foreground the contextual and sociotechnical nature of safety. Product safety standards are comparably focused on technical measures, and as such, corporate capture aside, these will never be sufficient to enforce comprehensive and meaningful AI safety management practices.

---

[31] https://corporateeurope.org/en/2025/01/bias-baked

[32] Leufer, D., & Hidvegi, F. (2023). The Pitfalls of the European Union's Risk-Based Approach to Digital Rulemaking. *UCLA L. Rev. Discourse*, *71*, 156.

[33] See footnote 31.

[34] We even go a step further to show that the deep linkages between (general-purpose) AI and planetary infrastructures and supply chains warrant seeing general-purpose AI as (part of) *social-ecological-technological systems*: Rakova, B., & Dobbe, R. (2023). Algorithms as Social-Ecological-Technological Systems: An Environmental Justice Lens on Algorithmic Audits. *2023 ACM Conference on Fairness, Accountability, and Transparency.* Full text: https://arxiv.org/pdf/2305.05733

Therefore, any narrow implementation of product safety standards mandated under the AI Act will require additional interventions, and organizations and authorities should prepare accordingly. The EU Code of Practice for General-Purpose AI Systems is one place to consider how and what more comprehensive interventions required to complement product safety standards may look like. System safety surely provides a fruitful blueprint, and many of the concerns and ideas surfaced in this response may carry over to the establishment of the code of practice.[35]

---

[35] Notably, the chair of the International AI Safety Report, Canadian Professor Yoshua Bengio, is also the chair of one of the working groups for the EU's General-Purpose AI Code of Practice, i.e. Working Group 3: Technical risk mitigation for systemic risk.

# 9 Why System Safety Can Help Shape "Public Interest AI"

The value of system safety concepts and methods goes beyond safety itself: these have been used to promote other public values. These include fairness, justice, access, privacy or security - all values that require a sociotechnical understanding of how these are realized across technical and non-technical aspects in society. Hence, if we take the history of system safety serious and translate the discipline's lessons to today, we may be able to address other values and concerns that are so vividly central to the implications of AI systems, including the protection of fundamental rights and environmental implications.

Put differently, we could say that big victories in ensuring the safety of other technological systems, including aviation, energy systems, road traffic and medical devices, meant that these went from potentially impactful to truly public interest technologies. Without system safety management, planes would have never become trusted and used by the broad public. Similarly, without sophisticated policies for traffic system safety, highways and cars would have never allowed for far distance high-speed ground traveling. And many analogies exist, be it for medical devices, energy systems or (bio)chemicals.

Making system safety a core tenet of public interest AI should give voice and power to those otherwise bearing the brunt of hazardous unaccountable technologies, be it as marginalized subjects or misunderstood professionals. It would inform clearer assignment of responsibilities and making transparent and accountable the normative foundations that are necessary to agree on what safe or unsafe really means. Furthermore, system safety practices would not just help minimize harm and resetting power to serve the public interest, but inform more rigorous and well-defined design of AI and digital technologies, as well as organizational and institutional innovations that are needed to translate public values to functional sociotechnical AI systems and sustainable democratic and rule of law practices.[36]

---

[36] Nouws, S. J. J., & Dobbe, R. I. J. (2024). The Rule of Law for Artificial Intelligence in Public Administration: A System Safety Perspective. In *Digital Governance: Confronting the challenges posed by Artificial Intelligence*. TMC Asser Press. https://surfdrive.surf.nl/files/index.php/s/gjt6Cg8RgxEVpWE

# 10 About the Sociotechnical AI Systems Lab

The Sociotechnical AI Systems Lab is located at Delft University of Technology in The Netherlands and develops system-theoretic principles and practices for the *integral analysis, design and governance of AI and algorithmic technologies in sociotechnical contexts*. Our work is motivated by the growing role of algorithms and AI in producing unsafe or unjust outcomes, often inflicted on marginalized communities. Our work is grounded in different sectors (in particular public services, energy systems and healthcare), engaging with key actors to understand and promote safety, and social and environmental justice in AI system design and governance.

Based on our transdisciplinary and empirical work, we develop and implement evidence-based understandings and system-theoretic approaches to AI system design and governance. Our holistic perspective, rooted in the long-standing tradition of *system safety and control theory,* provides a much-needed alternative to the dominance of technocentric approaches and their inability to address the systemic risks and opportunities of AI and algorithmic systems.

More information: https://surfdrive.surf.nl/files/index.php/s/4KkPCV3nYyDNeBT

# 11 About the Author

Roel Dobbe is a tenured Senior Assistant Professor in Technology, Policy & Management at Delft University of Technology.[37] He received a MSc in Systems & Control from Delft (2010) and a PhD in Electrical Engineering and Computer Sciences from UC Berkeley (2018). He was an inaugural postdoc at the AI Now. His research addresses the integration and implications of algorithmic technologies in societal infrastructure and democratic institutions, focusing on issues related to safety, sustainability and justice. His projects are situated in various domains, including energy systems, public administration, and healthcare. Roel's system-theoretic lens enables addressing the sociotechnical and political nature of algorithmic and artificial intelligence systems across analysis, engineering design and governance, with an aim to empower domain experts and affected communities. His results have informed various policy initiatives, including environmental assessments in the European AI Act as well as the development of the algorithm watchdog in The Netherlands.

---

[37] Personal website: https://www.tudelft.nl/staff/r.i.j.dobbe/